\documentclass[journal=jpccck,manuscript=article]{achemso}

\usepackage[version=3]{mhchem} % Formula subscripts using \ce{}
\usepackage{xcolor}
\usepackage[super]{natbib}
\usepackage{hyperref}
\author{Levi C. Felix}
\affiliation[State University of Campinas]
{Applied Physics Department, State University of Campinas, Campinas, SP, 13083-970, Brazil}
\alsoaffiliation[State University of Campinas]
{Center for Computational Engineering and Sciences, State University of Campinas, Campinas, SP, 13083-970, Brazil}
\author{Rushikesh Ambekar}
\affiliation[Indian Institute of Technology]
{Department of Metallurgical and Materials Engineering, Indian Institute of Technology Kharagpur, West Bengal, India}
\author{Raphael M. Tromer}
\affiliation[State University of Campinas]
{Applied Physics Department, State University of Campinas, Campinas, SP, 13083-970, Brazil}
\alsoaffiliation[State University of Campinas]
{Center for Computational Engineering and Sciences, State University of Campinas, Campinas, SP, 13083-970, Brazil}
\author{Cristiano F. Woellner}
\affiliation[Federal University of Paraná]
{Physics Department, Federal University of Paran\'a, Curitiba, PR,  Brazil}
\author{Varlei Rodrigues}
\affiliation[State University of Campinas]
{Applied Physics Department, State University of Campinas, Campinas, SP, 13083-970, Brazil}
\author{Pulickel M. Ajayan}
\affiliation[Rice University]
{Department of Materials Science and Nanoengineering, Rice University, Houston, TX, USA}
\author{Chandra S. Tiwary}
\affiliation[Indian Institute of Technology]
{Department of Metallurgical and Materials Engineering, Indian Institute of Technology Kharagpur, West Bengal, India}
\author{Douglas S. Galvao}
\email{galvao@ifi.unicamp.br}
\affiliation[State University of Campinas]
{Applied Physics Department, State University of Campinas, Campinas, SP, 13083-970, Brazil}
\alsoaffiliation[State University of Campinas]
{Center for Computational Engineering and Sciences, State University of Campinas, Campinas, SP, 13083-970, Brazil}
\title{From Pure Mathematics to Macroscale Applications:  The Genesis of Schwarzites}

\keywords{}

\begin{document}

%\begin{tocentry}

%\includegraphics[scale=1.0]{figs/graphical-abstract.pdf}
%\texttt{Schwarzites}
%Some journals require a graphical entry for the Table of Contents.
%This should be laid out ``print ready'' so that the sizing of the

%Inside the \texttt{tocentry} environment, the font used is Helvetica
%8\,pt, as required by \emph{Journal of the American Chemical

%The surrounding frame is 9\,cm by 3.5\,cm, which is the maximum
%permitted for  \emph{Journal of the American Chemical Society}
%graphical table of content entries. The box will not resize if the
%content is too big: instead it will overflow the edge of the box.

%This box and the associated title will always be printed on a
%separate page at the end of the document.

%\end{tocentry}

\begin{abstract}

  Schwarzites are porous (spongy-like) carbon allotropes with negative Gaussian curvatures. They were proposed by Mackay and Terrones inspired by the works of the German mathematician Hermann Schwarz on Triply-Periodic Minimal Surfaces (TPMS). 
  This review presents and discusses the history of schwarzites and their place among curved carbon nanomaterials. We summarized the main works on schwarzites available in the literature. We discuss their unique structural, electronic, thermal, and mechanical properties. Although the synthesis of carbon-based schwarzites remains elusive, the recent advances in the synthesis of zeolite-templates nanomaterials bring them closer to reality. Atomic-based models of schwarzites have been translated into macroscale ones that have been 3D printed. These 3D printed models have been exploited in many real-world applications, including water remediation and biomedical ones.

\end{abstract}

\section{Introduction}

The emergence of curved carbon nanostructures in the mid-1980s, such as fullerenes~\cite{kroto_1985} and nanotubes~\cite{iijima_1991}, have motivated the study and the advent of new ones~\cite{terrones_2003}. Inspired by the work of the German mathematician Hermann Schwarz (1843-1921), who did seminal works on the field of topology and geometry of periodic minimal surfaces~\cite{mackay_1985}, Mackay and Terrones proposed a new class of structures called schwarzites~\cite{mackay_1991}, after Schwarz.  Shwarzites are porous (spongy-like) three-fold coordinated carbon allotrope structures exhibiting negative Gaussian curvatures and with one-atom-thick pore-walls. They attracted much attention for their interesting mechanical~\cite{miller_2016,jung_2018,felix_2019,gong_2020}, electronic~\cite{huang_1993,valencia_2003,zhang_2010,lherbier_2014,weng_2015,owens_2016,gao_2017,tromer_2020}, thermal~\cite{pereira_2013,zhang_2017,zhang_2018,feng_2020}, structural~\cite{sajadi_2018,gaal_2021}, and many others properties~\cite{damascenoborges_2018,collins_2019,seok_2022}.

Due to their complex structural topologies, the synthesis of schwarzites represents a real challenge. First adopted as models for random graphene foams~\cite{townsend_1992,barborini_2002,benedek_2003,wu_2015}, the possibility of their synthesis through zeolite templating~\cite{nishihara_2009,nishihara_2018,braun_2018,boonyoung_2019} and/or molecular assembly into negative Gaussian-curvature sub-structures~\cite{ikemoto_2017,ikemoto_2018,ikemoto_2022}, have also been proposed, but their experimental realization remains elusive. However, recent advances in 3D print manufacturing technology have allowed the creation of 3D printed macroscale models of several Schwarzites structures~\cite{sajadi_2018}. These structures are lightweight and have high energy absorption capabilities~\cite{sajadi_2018,qin_2017}. These studies have inspired other works on 3D printed carbon-based lattices, such as tubulanes~\cite{sajadi_2019}, pentadiamond~\cite{felix_2022}, and zeolite-templated carbons~\cite{ambekar_2020}. Further works of 3D printed schwarzite have emerged in catalysis, wastewater treatment, building blocks, etc., exploiting engineering the interface with functional materials/nanomaterials for enhanced mechanical, absorption, and fracture-resistant properties~\cite{sajadi_2021}. In particular, using such porous structures in the remediation of microplastic-polluted waters has also shown promising results~\cite{gupta_2021}.

This review aims to contextualize the place of Schwarzites among curved carbon nanomaterials, summarize the main works on schwarzites available in the literature, and emphasize their potential in many daily-life applications. Since the property of negative Gaussian curvature is the one that mainly distinguishes them from other curved structures, the connection between topology and geometry of carbon materials will be briefly reviewed in the first section. Then, the stability of non-carbon schwarzites and their structural differences will be discussed. Separate sections will be dedicated to a summary of reported results in the literature on the mechanical, thermal, and adsorption properties. Its early use as models for graphitic random foams, alongside possible routes to their synthesis, will also be briefly reviewed. A summary of applications of 3D-printed macroscopic Schwarzites  is also discussed.

\section{Topology and Curved Carbon Nanomaterials}\label{sec:topologycarbon}

Carbon in its sp$^2$-hybridized state is found in graphite, graphene, and other planar allotropes. Such a well-defined hybridization state is attributed to the mirror symmetry of all $p_z$ orbitals, which does not distinguish where it is up or down the plane. However, the presence of a curvature breaks such symmetry by distorting these orbitals. Many curved carbon nanomaterials still possess three-fold coordination per atom, such as fullerenes~\cite{kroto_1985} and nanotubes~\cite{iijima_1991}, being also denominated graphitic structures. 

\begin{figure}
    \centering
    \includegraphics[width=\linewidth]{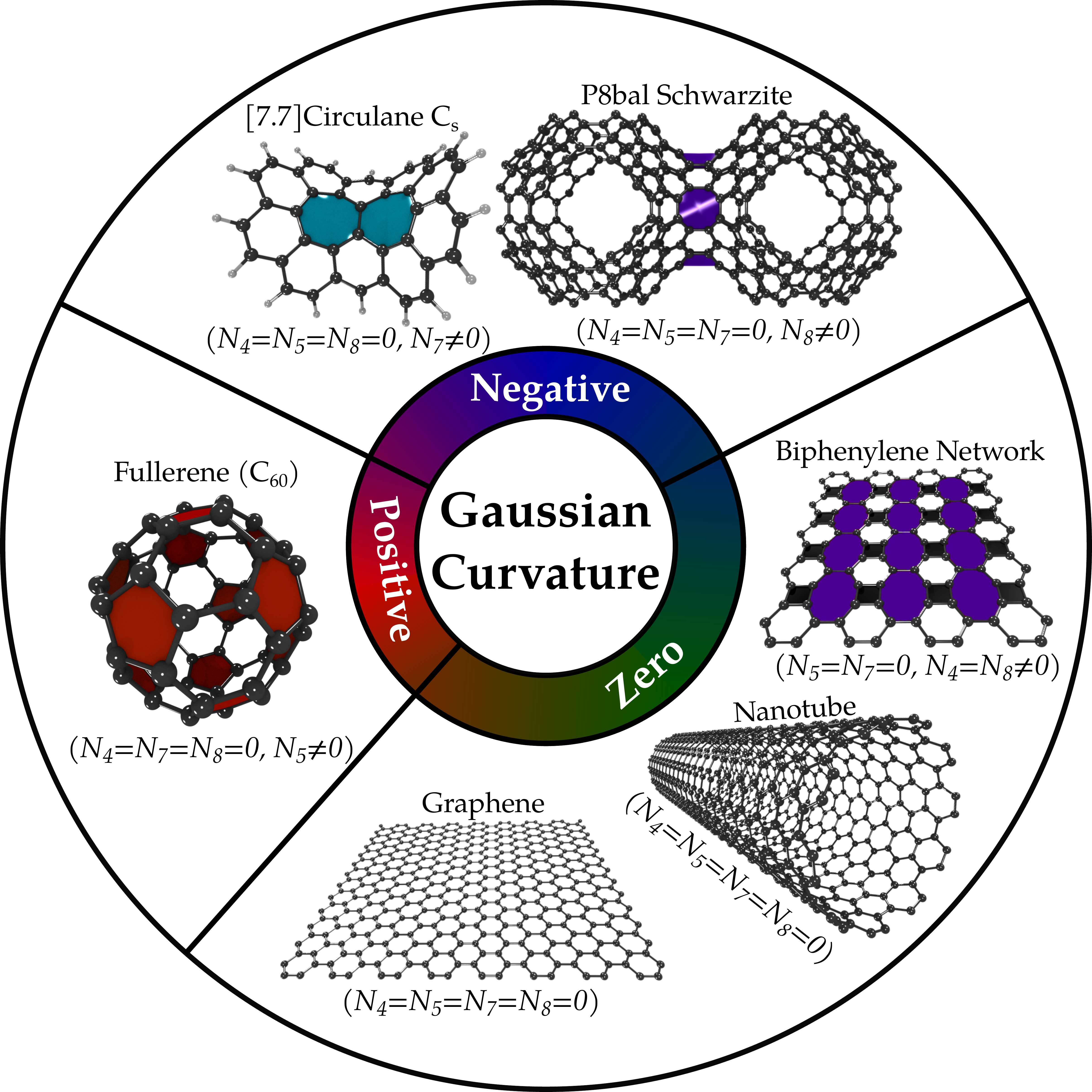}
    \caption{The Gauss-Bonnet-Euler theorem (Eq. \ref{eq:gauss-bonnet-euler}) establishes the relationship between the sign of Gaussian curvature and the ring distribution in curved structures. Six-membered rings do not contribute to the presence of curvature, as in graphene and nanotubes. The resultant surface is flat when squares and octagons are present in equal numbers, such as in biphenylene networks. A closed buckyball surface is obtained with the insertion of pentagonal rings. Negatively curved carbon structures can be built by introducing either seven- or eight-membered rings or a combination of other types of rings. Here, two examples of structures with negative Gaussian curvature are shown: a saddle-shaped molecule [7.7]Circulane C$_s$, and P8bal Schwarzite. Four-, five-, seven-, and eight-membered carbon rings are colored black, red, green, and purple, respectively.}
    \label{fig:carbon-topology}
\end{figure}

In most common planar allotropes of carbon, such as graphite and graphene, there are only six-membered carbon rings and no curvature. The existence and type of curvature can be uniquely determined by the presence of different types of rings that induce lattice distortions that are capable of creating curved (positive or negative) surfaces. These surfaces can be characterized by concepts such as Gaussian curvature, which connects geometry and topology by the Gauss-Bonnet-Euler formula
\begin{equation}\label{eq:gauss-bonnet-euler}
    \frac{3}{\pi}\int_S K dS = 2N_4 + N_5 - N_7 - 2N_8.
\end{equation}
Here, $K$ is the local Gaussian curvature, which, integrated over the surface $S$, gives the total Gaussian curvature. $N_k$ is the number of $k$-membered rings along $S$. Note that $N_6$ does not appear on Eq. \ref{eq:gauss-bonnet-euler} since hexagonal rings do not contribute to curvature. Even curved structures with only hexagons, which is the case of nanotubes~\cite{iijima_1991}, are characterized as having zero Gaussian curvature (see Figure \ref{fig:carbon-topology}). 

It is well known that the addition of five-membered rings on a graphene sheet can form stable structures. In fact, pentagonal rings alongside hexagons form fullerenes, such as $C_{60}$~\cite{kroto_1985}. According to Eq. \ref{eq:gauss-bonnet-euler}, this yields positive Gaussian curvature, as illustrated in Figure \ref{fig:carbon-topology}. Conversely, the presence of non-hexagonal rings may not result in curved structures if the right-hand side of Eq. \ref{eq:gauss-bonnet-euler} vanishes. This is the case of biphenylene carbon network~\cite{fan_2021}, with an equal number of four- and eight-membered rings along its plane, also shown in Figure \ref{fig:carbon-topology}.

\begin{figure}
    \centering
    \includegraphics[width=\linewidth]{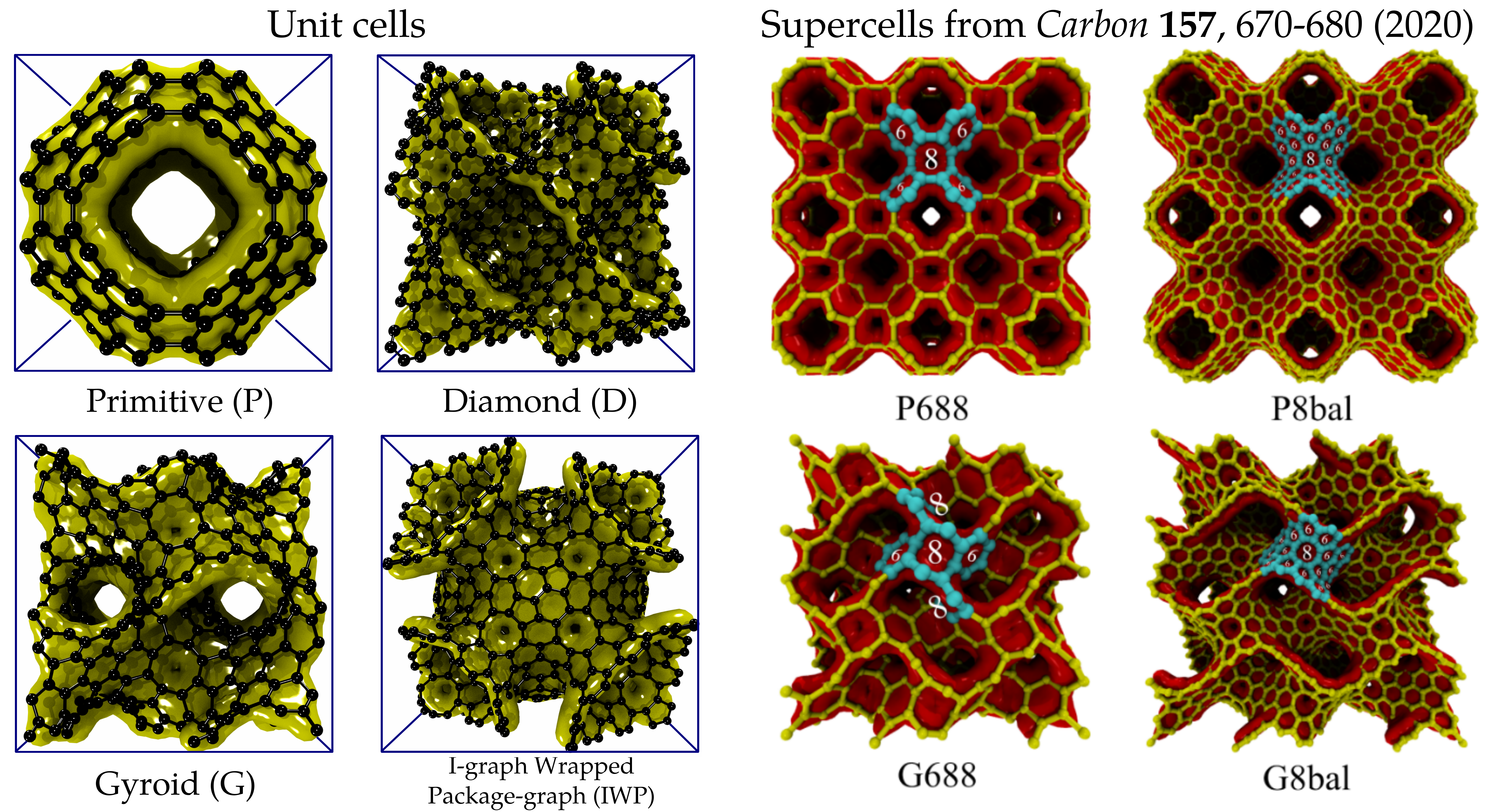}
    \caption{(Left) Unit cells of P, D, G, and IPWg families of TPMS/Schwarzites. (Right) Supercells for P and G families show the presence of eight-membered rings placed among several hexagons in blue. Reproduced with permission from reference \citenum{felix_2019}. Copyright 2020 Elsevier.}
    \label{fig:TPMS-schwarzites}
\end{figure}

Negatively curved structures can be built by combining different carbon rings in such a way that the right-hand side of Eq. \ref{eq:gauss-bonnet-euler} becomes negative. The simplest way of doing this is by introducing only heptagons or octagons. An example of the former includes saddle-shaped molecules such as a class called Circulanes~\cite{rickhaus_2017}. The first proposed schwarzites by Mackay and Terrones in 1991~\cite{mackay_1991} contained only octagonal rings as the geometrical curvature generators (shown in Figure \ref{fig:TPMS-schwarzites}). Whereas other Schwarzites containing five- and seven-membered rings (proposed by Lenosky and coworkers~\cite{lenosky_1992}) also possess negative curvature since they are placed in such a way that yields a negative total Gaussian curvature (see Eq. \ref{eq:gauss-bonnet-euler}). 

The solutions to the well-known mathematical problem of finding surfaces with minimal area and periodic along all three Cartesian directions are called Triply-Periodic Minimal Surfaces (TPMS). Pioneering in this field, Herman Schwarz~\cite{schwarz_1890}, found different families of TPMSs such as Primitive (P) and Diamond (D) (see Figure \ref{fig:TPMS-schwarzites}). Almost a century later, Alan Schoen~\cite{schoen_1970} found other families of surfaces that obey the properties required to be TPMSs, such as Gyroid (G) and I-graph Wrapped Package-graph (IWP). The first work that came up with the idea of decorating TPMS with graphitic (three-fold coordinated) atoms was that of Mackay and Terrones in 1991~\cite{mackay_1991}. In Figure \ref{fig:TPMS-schwarzites}, we can observe cubic unit cells of P, D, G, and IPWg families. Depending on the case, the use of supercells is necessary to obtain size-converged properties, as is the case for mechanical ones~\cite{felix_2019}, as shown in Figure \ref{fig:TPMS-schwarzites}.

Within a given schwarzite family, there are several different structures that differ mainly through their octagonal/hexagonal ring ratio, as can be seen in Figure \ref{fig:TPMS-schwarzites}. Structures with no hexagons separating two laterally neighbor octagons receive the suffix -688, as in P688 and G688, where the prefixes P and G refer to their families, such as primitive and gyroid, respectively. Structures with one hexagon between two consecutive octagons receive the suffix 8bal since eight-membered rings introduce curvature, and it balances both disconnected regions inside and outside pores in equal volumes. Note that other nomenclatures exist, such as Miller \textit{et al.}~\cite{miller_2016}, used 8-0 and 8-1, which substitutes 688 and 8bal, respectively.

\section{Mechanics of Schwarzites}\label{sec:mechanics}

\begin{figure}
    \centering
    \includegraphics[width=\linewidth]{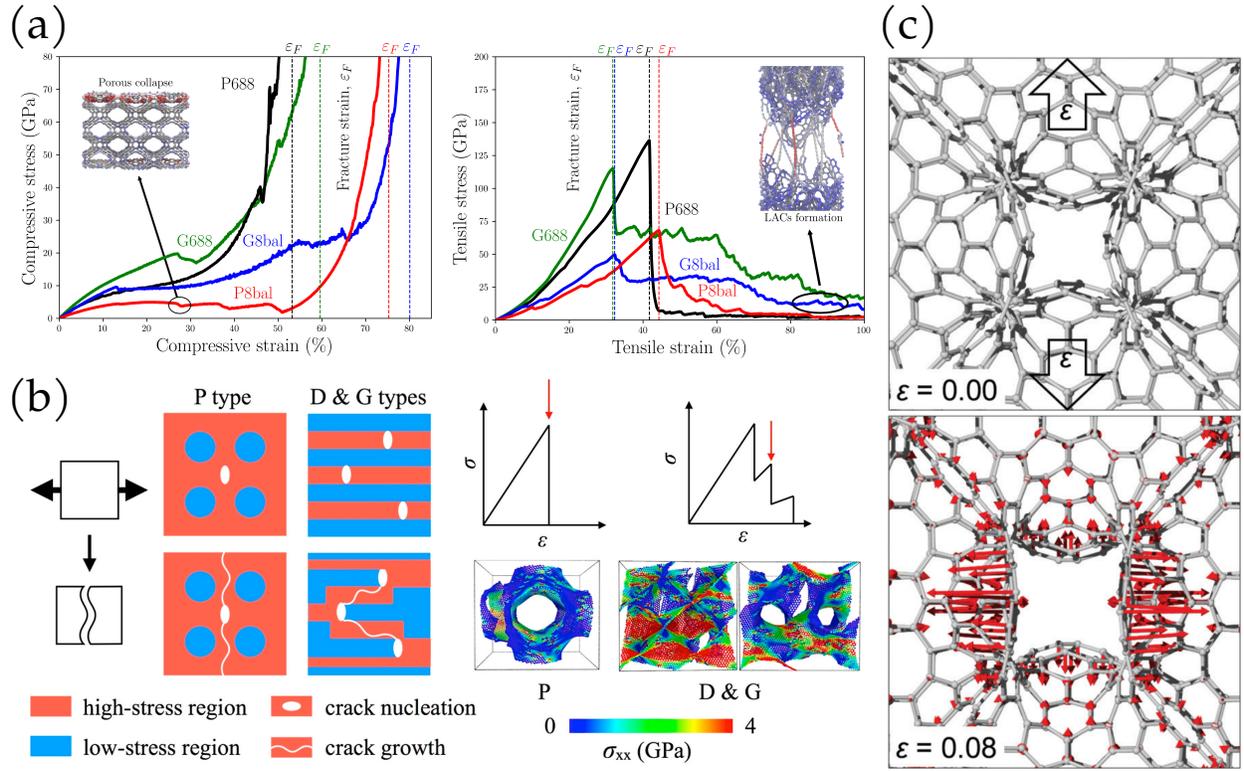}
    \caption{Mechanical properties of schwazites. (a) Stress-strain curves of schwarzites under compression and stretching. Pore collapses responsible for the presence of stress plateaus are shown on the inset, and the formation of LACs after rupture. Reproduced with permission of reference \citenum{felix_2019}. Copyright 2020 Elsevier. (b) Fracture patterns of different schwarzite families (types), with their characteristic stress-strain curve shapes and local atomic stress distributions. Reproduced with permission from reference \citenum{jung_2018}. Copyright 2018 American Chemical Society. (c) Auxetic behavior under stretching of IWPg schwarzite, where displacement vectors are represented as red arrows. Reproduced with permission from reference \citenum{gong_2020}. Copyright 2020 Royal Society of Chemistry.}
    \label{fig:mechanics}
\end{figure}

The larger the hexagonal/octagonal ring ratio, the larger the pores in a given schwarzite structure. There is a monotonic decrease in mechanical properties, such as in the Young and Bulk modulus values, as the pore size increases~\cite{miller_2016,felix_2019}. This is consistent with the fact that a larger pore size yields lower-density structures. On the other hand, these larger pore structures withstand larger deformation levels before ultimately breaking~\cite{felix_2019}. Under compressive loading, they possess a more well-defined stress plateau (Figure \ref{fig:mechanics}~(a)) where the appearance of kinks in this region is attributed to abrupt pore collapses (shown on the inset of Figure \ref{fig:mechanics}~(a)). Some structures can be compressed up to 80\% of strain before ultimately breaking.

Under stretching, fracture characteristics trends can be differentiated by their respective families instead of pore size. In the stress-strain curve of Figure \ref{fig:mechanics}~(a), G schwarzites tend to break before the P, and they are more susceptible to the formation of Linear Atomic Chains (LACs) (or Carbynes), as can be seen on the inset. The influence of LACs on the stress-strain response can be seen for strain levels after structural breaking as non-zero stress fluctuations. Carbyne is a very attractive material due to its high tensile strength~\cite{liu_2013} and tunable electronic properties~\cite{artyukhov_2014,latorre_2015}. The appearance of such LACs as a result of material fracture has been experimentally verified~\cite{jin_2009,chuvilin_2009,moras_2011,andrade_2015}. Although these chains are more energetically stable than other carbon allotropes, their high chemical reactivity (due to the presence of many unsaturated triple bonds) limits their fabrication on larger scales and, thus, the appearance of such LACs after schwarzite fractures would be hindered by this effect. As already shown in previous works~\cite{jung_2018}, the family type plays a significant role in the fracture pattern and, subsequently, in the stress-strain curve shape. Fracture patterns of P, G, and D families are shown in Figure \ref{fig:mechanics}~(b). For P Schwarzites, the fracture mechanism is dictated by the appearance of a single crack that propagates throughout the structure, leading to total rupture (as schematically shown in Figure \ref{fig:mechanics}~(b)), which occurs due to the sparse distribution of atomic stress. This behavior results in stress-strain curve shapes with a single peak followed by a drop to zero stress. The presence of regions with high atomic stress separated by small stress-value regions, shown for D and G types, favors the nucleation of many cracks prior to total rupture. When the first crack is created, the stress is partially released and then restarts to increase until the fracture propagates to the nearest crack, where the total stress decreases again, yielding a sawlike-shape stress-strain curve (shown in Figure \ref{fig:mechanics}~(b)).

Negative Poisson's ratio (also known as auxetic behavior) has also been reported for some schwarzite structures~\cite{felix_2019,gong_2020}. In Figure \ref{fig:mechanics}~(c), it is shown that the IWPg structure expands laterally while being uniaxially stretched. This occurs due to re-entrant topology~\cite{ashby_2006} present in the crystal structure of IWP schwarzites, which holds for both compression and tension~\cite{gong_2020}.

\section{Thermal Conduction and Hybridization of Phonon Modes}\label{sec:thermal}

\begin{figure}
    \centering
    \includegraphics[width=\linewidth]{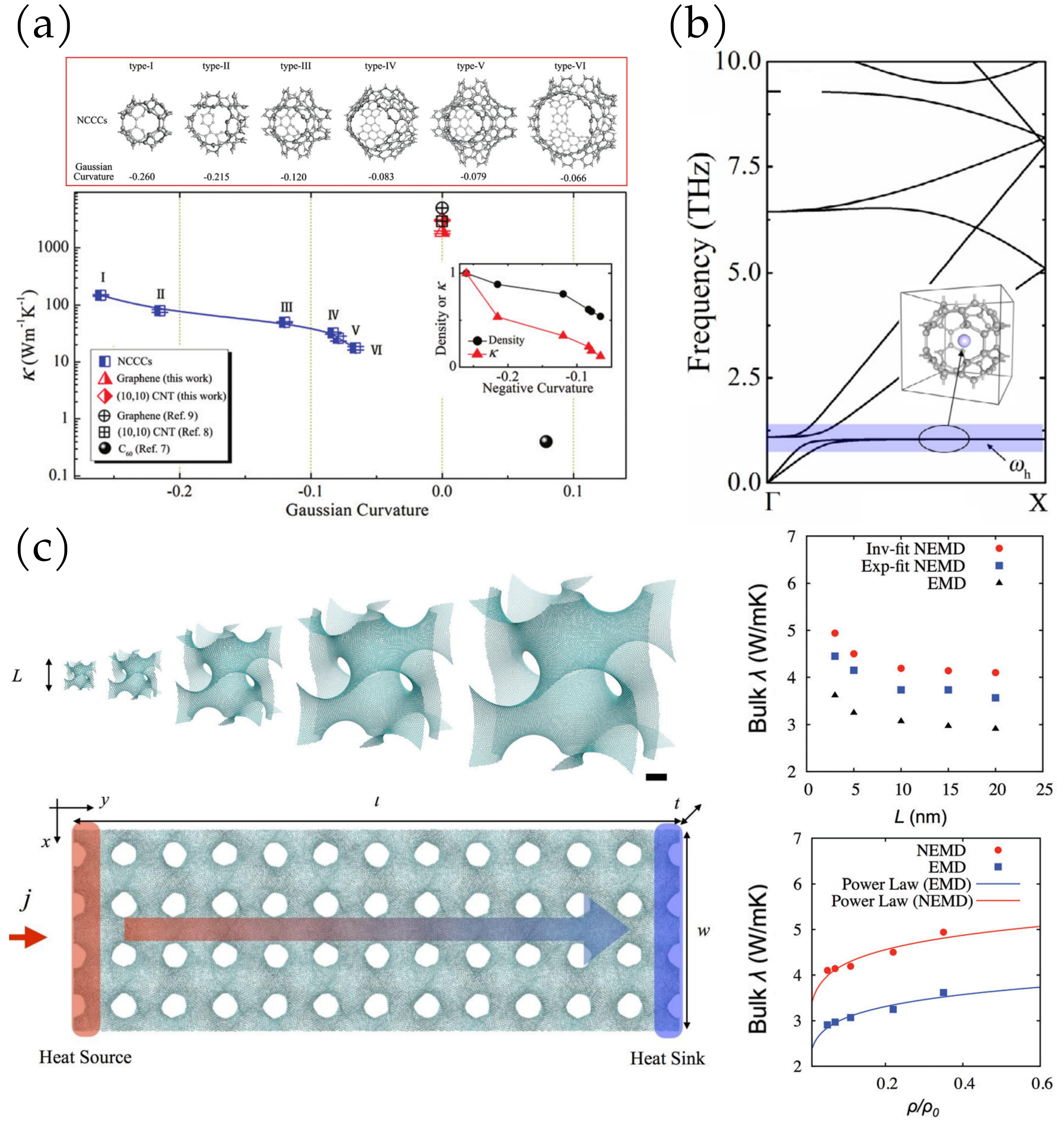}
    \caption{Thermal transport in schwarzites. (a) Decrease of thermal conductivity caused by the presence of curvature for P schwarzites compared to other carbon allotropes. Reproduced with permission from reference \citenum{zhang_2017}. Copyright 2017 Royal Society of Chemistry. (b) Hybridization of acoustic phonon branches in P688 schwarzite as a guest-host system. Reproduced with permission from reference \citenum{zhang_2018}. Copyright 2018 Elsevier. (c) Non-equilibrium molecular dynamics (NEMD) calculations of the thermal conductivity of G schwarzites as a function of the unit-cell size. Reproduced with permission from reference \citenum{jung_2017}. Copyright 2017 Royal Society of Chemistry.}
    \label{fig:thermal}
\end{figure}

Zero curvature sp$^2$ carbon nanomaterials such as graphene and carbon nanotubes possess extremely high lattice thermal conductivity of $\sim 1015$ and $\sim 1700$~Wm$^{-1}$K$^{-1}$, respectively~\cite{pereira_2013}. However, the presence of surface curvature dramatically suppresses thermal transport, as reported for several schwarzites~\cite{pereira_2013,zhang_2017} (called Negatively Curved Carbon Crystals (NCCCs)). Figure \ref{fig:thermal}~(a) illustrates this decrease in thermal conductivity as a function of the nominal value of Gaussian curvature for several P schwarzites. It can be seen that the type number increases with the relative number of hexagons in the unit cell (similar to the idea illustrated in Figure \ref{fig:TPMS-schwarzites}), which translates into structures with lower absolute values of curvature. A monotonic decrease in conductivity is observed up to the case where the curvature is zero, which is the case of graphene and nanotubes. More dense schwarzite models, such as Type I in Figure \ref{fig:thermal}~(a) (P688), provide better ways of sound propagation that are reflected in higher group velocities of acoustic modes in the phonon spectrum~\cite{zhang_2017}. This is a possible mechanism that might explain the thermal conductivity decrease upon increasing Gaussian curvature values (lowering density, shown in the inset of Figure \ref{fig:thermal}~(a)).

The porous structure, allied with the low thermal conductivity of schwarzites, makes them suitable candidates for building guest-host systems for thermoelectric applications. In such systems, chemically inert atoms (guests) are placed inside the available pores of a crystalline framework (host) to induce modifications in the vibrational modes of the lattice. An illustration of such configuration can be seen on the inset of Figure \ref{fig:thermal}~(b). There, a generic inert atom interacts with the cage of a P688 schwarzite through Lennard-Jones potential. The resultant effect is the anti-crossing bands at the blue-hatched region of the phonon dispersion, which is attributed to the coupling between longitudinal and transversal acoustic modes (mainly responsible for phonon transport). This further reduces thermal conductivity by decreasing the slopes of acoustic branches~\cite{zhang_2018}. 

The results on thermal transport presented so far were obtained by the Green-Kubo formalism, where transport coefficients are obtained from fluctuations of a suitable quantity. For instance, thermal conductivity is the autocorrelation of the heat flux. These results are collected from molecular dynamics simulations of the system at dynamic equilibrium, which is denoted by Equilibrium Molecular Dynamics (EMD). As we know, thermal transport is an out-of-equilibrium phenomenon that requires the presence of a temperature gradient (absent in EMD). Thus, it is also common to calculate thermal conductivity from simulations where parts of the system are kept at constant temperatures (heat source and cold sink), also known as Non-Equilibrium Molecular Dynamics (NEMD). Figure \ref{fig:thermal}~(c) shows an example of such a configuration, where a G schwarzite supercell is subjected to a temperature gradient. Using different unit-cell sizes (shown in Figure \ref{fig:thermal}~(c)), size-dependence of thermal conductivity is obtained as the ratio between the induced heat flux and the corresponding chosen thermal gradient, according to heat Fourier's law. Similar to the case of EMD results for P schwarzites~\cite{zhang_2017}, thermal conductivity is reduced by increasing the unit cell size (decreasing the density), as shown in Figure \ref{fig:thermal}~(c).

\section{Schwarzites as Frameworks for Gas Adsorption}\label{sec:adsorption}

\begin{figure}
    \centering
    \includegraphics[width=\linewidth]{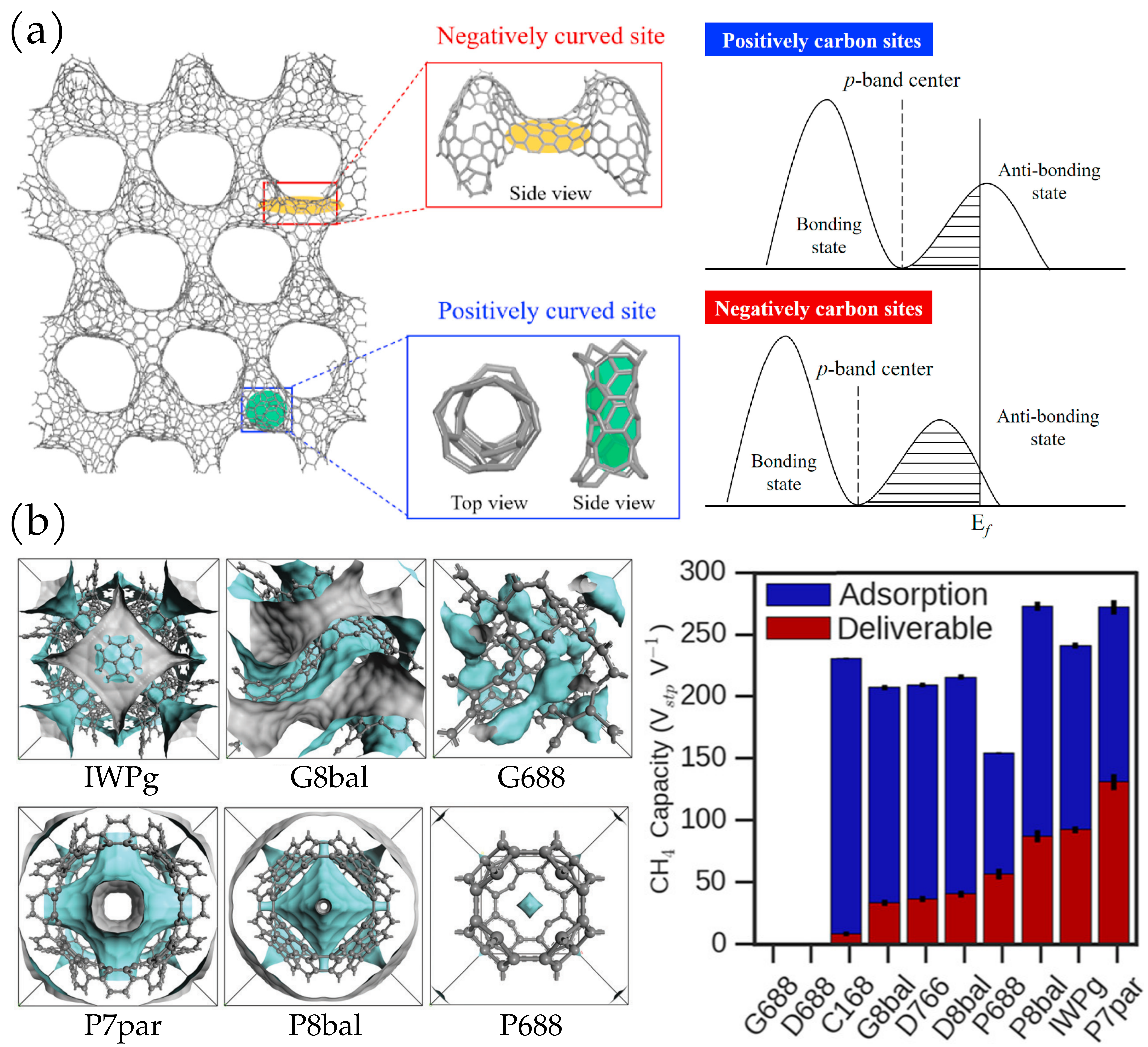}
    \caption{Gas adsorption in schwarzites cages. (a) Hydrogen storage in positively and negatively curved sites alongside their relative p-band center positions. Reproduced with permission from reference \citenum{seok_2022}. Copyright 2022 Elsevier. (b) Methane adsorption for different schwarzite structures. Reproduced with permission from reference \citenum{collins_2019}. Copyright 2019 American Chemical Society.}
    \label{fig:adsorption}
\end{figure}

The search for nanoporous materials to capture and/or store natural molecular gases is still an ongoing research topic that attracts the scientific community's interest due to the goal of going green. In particular, H$_2$ and CO$_2$ storage are the most investigated gases due to their relevance in hydrogen fuel technology and carbon dioxide emission decrease~\cite{damascenoborges_2018}. A systematic understanding of how Gaussian curvature influences adsorption properties might be crucial to design strategies to optimize molecule capture/storage performance. It has been reported that the presence of regions with local curvature both positive and negative (see Figure \ref{fig:adsorption}~(a)) enhances the binding energy of some specific chemisorption sites for molecular hydrogen~\cite{seok_2022}. Although schwarzites are known for their negative Gaussian curvature, this value refers to the total value averaged throughout their surfaces. In contrast, locally, the curvature values can be positive and negative for some regions. Namely, positively curved sites are more favorable to the formation of chemisorption because the anti-bonding state of their $p$-bands is less partially filled to those corresponding to the negatively curved sites~\cite{seok_2022}, as shown in Figure \ref{fig:adsorption}~(a).

Another molecular gas whose storage is a target of intense research is methane. Its use in vehicular applications as natural gas fuel has been focused on as an alternative to fossil fuels, significantly contributing to the greenhouse effect. Several porous materials~\cite{wu_2016,mercado_2018} have been investigated regarding their methane storage capabilities. The porous structure of schwarzites has also gained attention from the scientific community in this sense~\cite{damascenoborges_2017,collins_2019}. The smallest schwarzites from 4 different families have been studied (Figure \ref{fig:adsorption}~(b)), with large-pore structures having higher adsorption capabilities, as shown in the histogram of Figure \ref{fig:adsorption}~(b). This occurs due to the well-known trend of larger pores, providing a larger surface area for a greater adsorption capacity.

\section{The Stability of Non-carbon Schwarzites}\label{sec:differentelements}

As it is well-known for the cases of diamond and graphene crystalline structures, a version of schwarzites made of other chemical elements has already been proposed in several works where the carbon atoms are replaced by boron nitrde~\cite{gao_2017}, silicon~\cite{laviolette_2000,zhang_2010}, and, most recently, germanium~\cite{tromer_2020}. The stronger sp$^3$ character of silicon (Si) and germanium (Ge) bonding produces significant structural differences in all carbon nanostructured analogs. For instance, silicene and germanene membranes differ from purely flat graphene by an out-of-plane buckling of the atomic structure~\cite{ni_2012,ohare_2012,vogt_2012,dvila_2014}. This 
occurs due to the presence of the pseudo-Jahn-Teller effect~\cite{hobey_1965,bersuker_2013,perim_2014}.

\begin{figure}
    \centering
    \includegraphics[width=\linewidth]{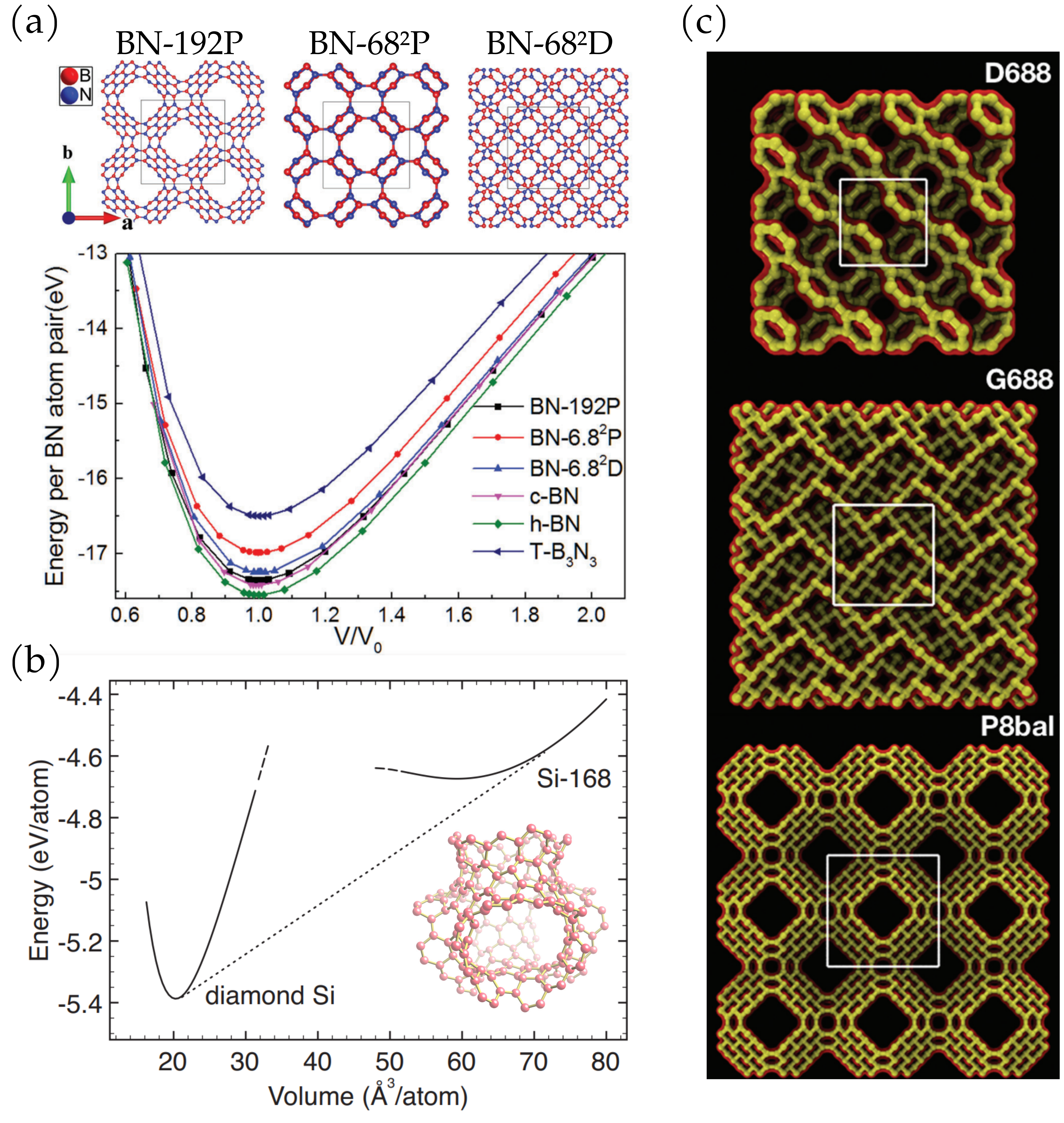}
    \caption{Non-carbon schwarzites. (a) P8bal, P688, and D688 BN schwarzites with their energy-volume curves. Reproduced with permission from reference \citenum{gao_2017}. Copyright 2017 Royal Society of Chemistry. (b) Energy as a function volume per atom of an \textit{fcc} Si-168 schwarzite. Reproduced with permission from reference \citenum{zhang_2010}. Copyright 2010 American Institute of Physics. (c) Buckled structures of D688, G688, and P8bal Ge schwarzites with their unit cell highlighted on white squares. Reproduced with permission from reference \citenum{tromer_2020}. Copyright 2020 Royal Society of Chemistry.}
    \label{fig:other-elements}
\end{figure}

Similar to flat hexagonal boron nitride (BN), each boron atom only has nitrogen atoms as nearest neighbors in BN schwarzites, as seen in Figure \ref{fig:other-elements}~(a). The presence of a single minimum indicates the structural stability of all BN structures reported therein, where although they are metastable than the usual hexagonal BN (hBN) and cubic BN, they are more stable than previously reported T-B$_3$N$_3$. P8bal BN schwarzite is the most stable among all three structures. Relative energetic stability between a silicon schwarzite and the more usual silicon in a diamond structure~\cite{zhang_2010} is of the same order of magnitude as those of BN (Figure \ref{fig:other-elements}~(b). Recent studies considering germanium schwarzites showed pronounced local structural distortions (compared to carbon) due to the pseudo-Jahn-Teller effect for the largest structures~\cite{tromer_2020} while showing relative stability compared to other germanium structures.

\section{Possible Roadmaps to Schwarzite Synthesis}\label{sec:synthesis}

\begin{figure}
    \centering
    \includegraphics[width=\linewidth]{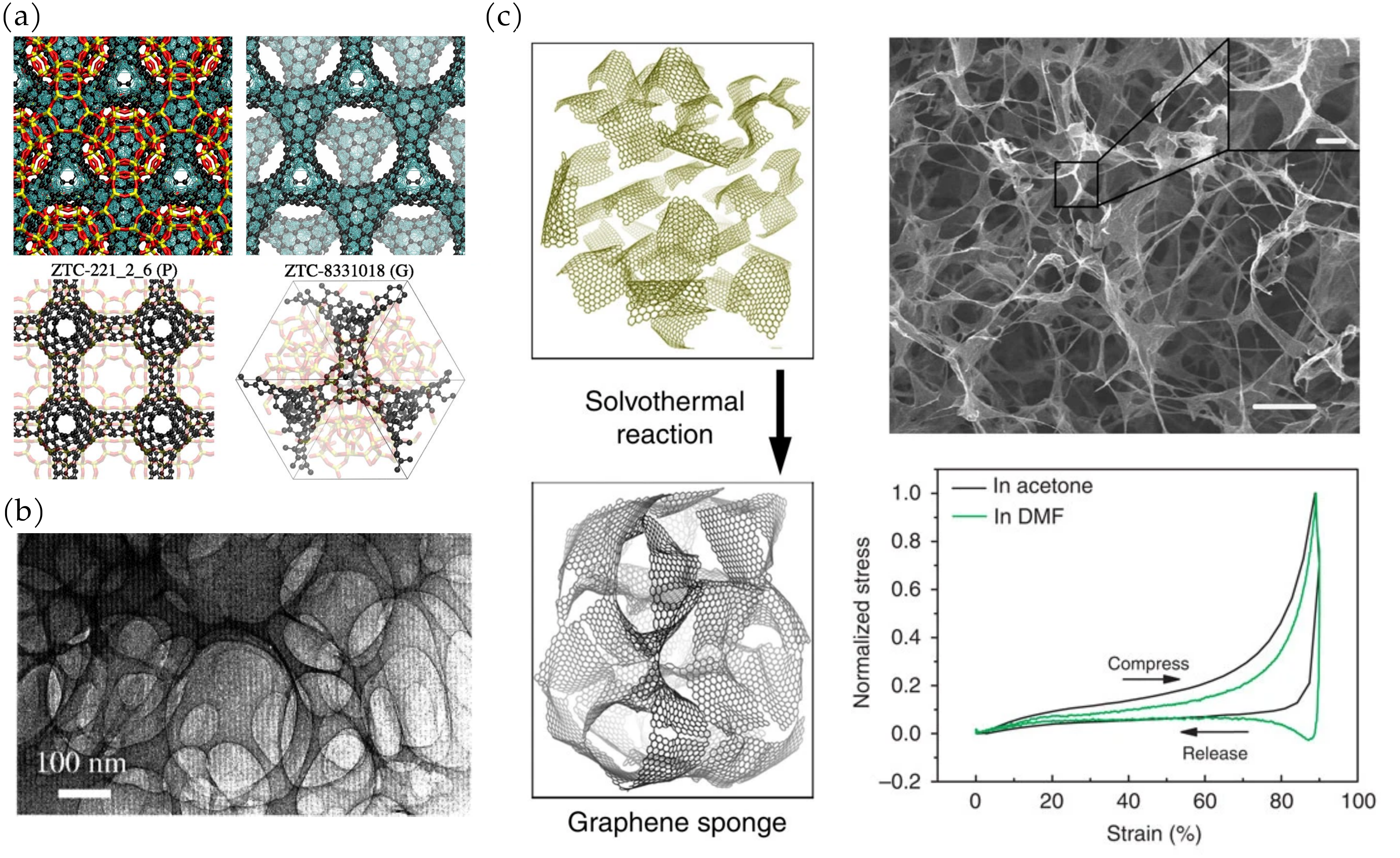}
    \caption{Roadmap to schwarzites synthesis. (a) Illustration of \textit{in silico} schwarzites generation by Zeolite-Templating Carbon alongside two examples of produced structures. Reproduced with permission from reference \citenum{braun_2018}. Copyright 2018 National Academy of Sciences of the United States of America. (b) Negatively curved spongy carbon modeled by D schwarzites. Reproduced with permission from reference \citenum{benedek_2003}. Copyright 2003 Elsevier. (c) Spongy carbon synthesis illustration, scanning electron microscope image, and stress cycle. Reproduced with permission from reference \citenum{wu_2015}. Copyright 2015 Nature.}
    \label{fig:synthesis}
\end{figure}

Although a well-controlled synthesis of schwarzites is yet to be achieved, numerous efforts have been reported in the literature~\cite{nishihara_2009,nishihara_2018,boonyoung_2019,braun_2018}. Especially, \textit{in silico} approaches have gained attention in the last few years~\cite{braun_2018} regarding the production of zeolite-templated carbon (ZTC). Several ZTCs have geometry similar to schwarzites, including the generation of different families, such as G and D~\cite{braun_2018} (see Figure \ref{fig:synthesis}~(a)). The claimed experimental feasibility of such a protocol is mainly based on the idea of removing the zeolite framework with acidic compounds while maintaining the whole carbon structure at the end of the process.

Although experimental synthesis of crystalline schwarzites remains elusive, their porous character motivated many approaches~\cite{townsend_1992,benedek_2003,wu_2015,zhu_2022} to use them as models for graphitic random nanofoams. Their predominant sp$^2$ character provides them with the characteristic yielding and compliance of cell walls of schwarzites. Namely, they can also be compressed to a high degree of deformations before ultimately breaking~\cite{benedek_2003,wu_2015}. Figure \ref{fig:synthesis}~(b) shows an example of a graphitic nanofoam image from a scanning tunneling microscope synthesized by a pulsed microplasma cluster source in the presence of a metal-organic catalyst, and it can be deposited as a film by supersonic cluster beam deposition. This form of carbon is characterized by interconnected thin layers forming a spongy structure with meso- and macroporosity. It consists of a robust, multiply-connected graphene sheet fully covalent along the three dimensions~\cite{benedek_2003}. Assembly of ultra-large-area graphene sheets in solution via solvothermal reactions (Figure \ref{fig:synthesis}~(c)) is another route that has been applied to fabricate such carbon nanofoams~\cite{wu_2015}. These graphene sponges shown in Figure \ref{fig:synthesis}~(c), with densities similar to air, display near-zero Poisson’s ratios along all directions and are largely strain-independent during reversible compression to giant strains. At the same time, they function as enthalpic rubbers, which can recover up to 98\% compression in air and 90\% in liquids and operate between -196 and 900$^o$C. Furthermore, these sponges provide reversible liquid absorption for hundreds of cycles and then discharge it within seconds while still providing an effective near-zero Poisson’s ratio~\cite{wu_2015}. 

\section{Macroscopic 3D-Printed Schwarzites}\label{sec:3dprinting}

\begin{figure}
    \centering
    \includegraphics[width=1.0\linewidth]{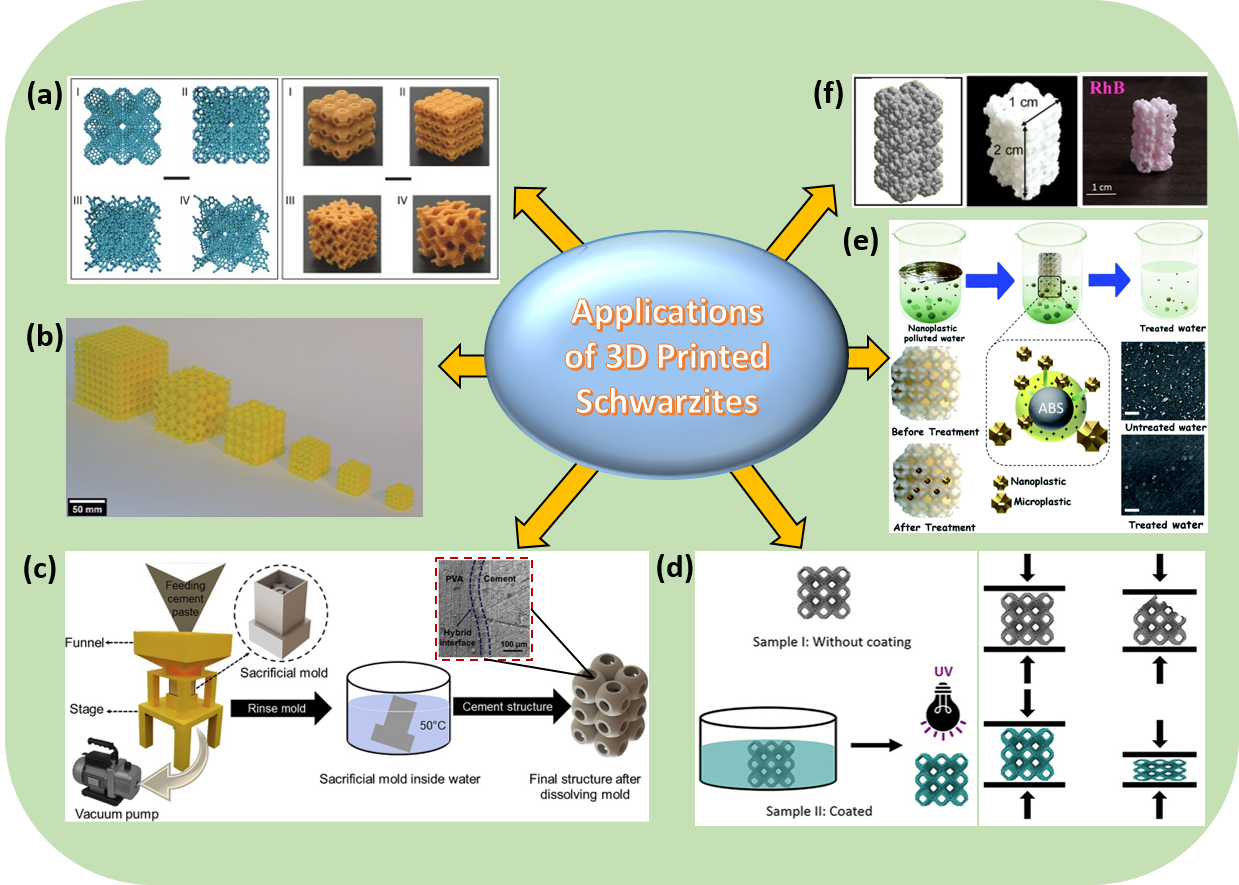}
    \caption{ 3D-printed schwarzites. (a) Atomic models and 3D-printed schwarzite structures made of of PLA polymer. Reproduced with permission from reference \citenum{sajadi_2018}. Copyright 2018 Wiley. (b) 3D-printed schwarzites with different sizes, showing the same scale factor as atomic models. Reproduced with permission from reference \citenum{gaal_2021}. Copyright 2022 Wiley. (c) The fabrication method of cementitious schwarzite. Reproduced with permission from reference \citenum{sajadi_2021}. Copyright 2021 Elsevier. (d) Conformal coating strategy of epoxy-coated ceramic schwarzite and schematic of mechanical performance with and without coating. Reproduced with permission from reference \citenum{sajadi_2021b}. Copyright 2021 AAAS.  (e) Development of a schwarzite-based moving bed 3D printed water treatment system for nanoplastic remediation. Reproduced with permission from reference \citenum{gupta_2021}. Copyright 2021 Royal Society of Chemistry. (f) before and after 3D printed schwarzite and interaction of 3D printed schwarzite with Rhodamine B dye. Reproduced with permission from reference \citenum{kumbhakar_2021}. Copyright 2021 Elsevier.}
    \label{fig:3d-printing}
\end{figure}

The challenge imposed by synthesizing schwarzites at an atomic scale prevents exploring their exciting properties, as mentioned in this review. This shortcoming has led to some pioneer works to consider the development of schwarzite topologies at a macroscale with the use of additive manufacturing techniques~\cite{sajadi_2018,qin_2017,sajadi_2019,gaal_2021,ambekar_2020,felix_2022}. Sajadi \textit{et al.}~\cite{sajadi_2018} explored a multi-scale geometrical design of 3D-printed schwarzites (as seen in Figure \ref{fig:3d-printing}~(a)), where an interpolated surface generated from the atomic positions serves as the basis for 3D printing of polymeric structures which are subsequently mechanically compressed. They have observed that the mechanical characteristics of schwarzites structures are similar regardless of the material and length scale difference. This is a signature of the distinct surface topology of each individual structure. The deformation study reveals that Schwarzite structures have unique layered deformation and uniform distribution of stress concentration. Such topology-dependent properties are more pronounced when the scale factor is the same between both scales, as was done by Ga\'al \textit{et al.}~\cite{gaal_2021}. Figure \ref{fig:3d-printing}~(b) shows 3D-printed schwarzites with the same size scale relative to each individual structure. Also, some stress collapses are observed in both length scales at roughly the same strain levels. This work also considered different additive manufacturing polymers such as Poly-Lactic Acid (PLA) and Acrylonitrile Butadiene Styrene (ABS). PLA was found to be more well-behaved during mechanical compression and withstand more significant strains without breaking~\cite{gaal_2021}. Another direction is to hierarchically combine different schwarzite topologies to increase the possibilities for their exceptional mechanical response. Some recent strategies include the introduction of sp fragments into their cage-structures~\cite{oliveira_2022} (named as Schwarzynes) and the generation of polycrystalline domains with tunable response~\cite{ambekar_2021}. A recent structural application of schwarzites was explored in damping enhancement and vibration isolation. Due to its excellent mechanical strength, schwarzite exhibits an 80\% higher damping ratio than the solid base material. Also, it can withstand frequency shifts lower than excitation, resulting in vibration isolation. In this way, vibration isolation and increased energy dissipation result in lower force transmissibility~\cite{herkal_2023}.

Researchers have verified the feasibility of schwarzite in building and construction areas, as the present building and construction materials possess poor fracture strain. The PVA-coated schwarzite cement structure was developed via 3D printing and investment casting. 3D printing technique was utilized for the fabrication of a complex schwarzites mold structure using polyvinyl alcohol, and then cement was cast in it. Afterward, PVA mold was dissolved in a suitable solvent, resulting in PVA-coated schwarzite cement structure (Figure \ref{fig:3d-printing}~(c)). The specific energy absorption and fracture strain of cementitious schwarzite structure is two times and 17 times, respectively, greater than Cementitious solid structure~\cite{sajadi_2021}. Similarly, epoxy-coated ceramic Schwarzite structures were developed via 3D printing and conformal coating (Figure \ref{fig:3d-printing}~(d)) to enhance the structural stability of ceramic structures at higher strains. Synergistic improvement in specific energy absorption (77 times) and fracture strain (13 times) was observed compared to the uncoated ceramic schwarzite structure. They have observed that epoxy coating helps delay crack propagation~\cite{sajadi_2021b}. 

Due to their high surface area, schwarzites can be used for carrier applications in different fields. 3D printed schwarzite carriers were used for the remediation of nanoplastics (Figure \ref{fig:3d-printing}~(e)). The selection of material is based on the molar interaction constant; therefore, ABS-made schwarzite carriers were used for the removal of polycarbonate due to electrostatic interaction between them~\cite{gupta_2021}. Similarly, catalysis is another application where surface area plays an important role. Dye degradation capability and water cleaning efficiency of ZnO-decorated 3D-printed schwarzite structures were studied. The accelerated removal efficiency ($\sim 80\%$) was experienced with ZnO decorated schwarzite carrier against Rhodamine B (Figure \ref{fig:3d-printing}~(f)) dye due to the high surface area of schwarzite and catalytically active material (ZnO)~\cite{kumbhakar_2021}. Bone regeneration is another field that could benefit from the structural robustness of TPMS/Schwarzites topology~\cite{hoel_2023,hayashi_2023,pugliese_2023}. Their high surface area also enhances the energy harvesting from rain droplets~\cite{kumbhakar_2022} and CO$_2$ capture~\cite{ellebracht_2023} along their surface, proving their multifunctional capabilities.

\section{Conclusions and future work}

Schwarzites represent an interesting case in science, where the original concept was based on pure mathematics, and then the topological concepts were translated into carbon allotropes with negative curvature. Atomic-based models were proposed and investigated for their structural, electronic, mechanical and transport properties. The atomic models were also translated into macroscale ones that were 3D printed. Remarkably, some mechanical properties were found to be material and scale-independent, indicating that the topology determines most of the mechanical behavior. The negative Gaussian curvature of schwarzites is a unique topology amongst the different carbon allotropes. The appealing mechanical properties of schwarzites are due to complex topology and uniform stress distribution. Thermal conductivity can be tuned by varying unit cells of schwarzites. Gas adsorption in schwarzite cages can also be tuned by systematically designing the unit cell size. Real-world applications of 3D printed schwarzites have been already demonstrated for water remediation and biomedical applications. Recently, researchers have built a multi-layered schwarzite architecture named Petal Schwarzite, which shows unique mechanical properties~\cite{bastos_2023,ambekar_2023}. It is possible to take advantage of the excellent mechanical properties and high surface area of schwarzite in various fields by developing multifunctional schwarzite structures for energy harvesting, thermal management systems, sound absorbers, sensors, microfluidic systems, water filters, etc. The recent synthetic advances in zeolite-templated materials make schwarzities closer to being experimentally realized. We hope the present work will stimulate further studies on these remarkable structures.

\begin{acknowledgement}

The authors acknowledge the Center for Computing in Engineering and Sciences at Unicamp for financial support through the FAPESP/CEPID Grants \#2013/08293-7 and \#2018/11352-7.

%Rappe, A. K.; Casewit, C. J.; Colwell, K. S.; Goddard, W. A., III;Skiff, W. M. UFF, a full periodic table force field for molecularmechanics and molecular dynamics simulations.J. Am. Chem. Soc.1992,114, 10024−10035.

\end{acknowledgement}


\begin{suppinfo}

\end{suppinfo}

\bibliography{achemso-demo}

\end{document}